\documentclass[aps,a4paper,pre,floatfix,showkeys,twocolumn]{revtex4}
\usepackage{graphicx}
\usepackage{amsmath}
\usepackage{amssymb}
\usepackage{wasysym}
\usepackage{fdsymbol}
\usepackage{setspace}
\usepackage{color}
\usepackage[dvipsnames]{xcolor}

\begin{document}


\title{Polymer-induced microcolony compaction in early biofilms:\\ a computer simulation study}

\author{Francisco Javier Lobo-Cabrera$^1$}

\author{Alessandro Patti$^2$}

\author{Fernando Govantes$^3$}

\author{Alejandro Cuetos$^1$}
\email{acuemen@upo.es}

\affiliation{$^1$Department of Physical, Chemical and Natural Systems, Pablo de Olavide University, 41013 Sevilla, Spain}
\affiliation{$^2$Department of Chemical Engineering and Analytical Science, The University of Manchester, Manchester, M13 9PL, UK}
\affiliation{$^3$Centro Andaluz de Biolog\'{\i}a del Desarrollo (Universidad Pablo de Olavide, Consejo Superior de 
	Investigaciones Cient\'{\i}ficas y Junta de Andaluc\'{\i}a) and Departamento de Biolog\'ia Molecular e 
	Ingenier\'ia Bioqu\'imica, Universidad Pablo de Olavide}

\begin{abstract}

Microscopic organisms, such as bacteria, have the ability of colonizing surfaces and developing biofilms that can determine diseases and infections. Most bacteria secrete a significant amount of extracellular polymer substances that are relevant for biofilm stabilization and growth. In this work, we apply computer simulation and perform experiments to investigate the impact of polymer size and concentration on early biofilm formation and growth. We observe as bacterial cells formed loose, disorganized clusters whenever the effect of diffusion exceeded that of cell growth and division. Addition of model polymeric molecules induced particle self-assembly and  aggregation to form compact clusters in a polymer size- and concentration-dependent fashion. We also find that large polymer size or concentration lead to the development of intriguing stripe-like and dendritic colonies. The results obtained by Brownian dynamic simulation closely resemble the morphologies that we experimentally observe in biofilms of a Pseudomonas Putida strain with added polymers. The analysis of the Brownian dynamic simulation results suggests the existence of a threshold polymer concentration that distinguishes between two growth regimes. Below this threshold, the main force driving polymer-induced compaction is hindrance of bacterial cell diffusion, while collective effects play a minor role. Above this threshold, especially for large polymers, polymer-induced compaction is a collective phenomenon driven by depletion forces. Well above this concentration threshold, severely limited diffusion drives the formation of filaments and dendritic colonies.

\end{abstract}


\maketitle

\section{Introduction}
	
Bacterial biofilm growth results from the lengthening and division of individual bacteria attached to a surface and leads to the formation of flat, surface-bound microcolonies \cite{COS95}.  The latter eventually evolve into complex three-dimensional biofilm colonies involving large numbers of individual cells and displaying complex  shapes and structures. In its early stages, biofilm formation can be regarded as a competition between passive cell diffusion and active lengthening and division \cite{ACE18}. If bacterial multiplication predominates over diffusion, compact structures displaying a degree of orientational and positional order are formed. In contrast, loose swarm-like ensembles of scattered bacterial cells are usually found when diffusion prevails over lengthening and division \cite{ACE18}. Development of bacterial biofilms is a process determined by nutrient consumption. Despite its persistent out--of--equilibrium character, the mechanism of biofilm formation suggests an parallelism with the early stage of liquid-crystal nucleation in a metastable isotropic fluid, where nematic domains (or tactoids) are generally observed \cite{CUE08, JAM15}. 

A key factor in biofilm development is the presence of extracellular polymer substances (EPS) in the environment surrounding the bacteria. A large fraction of EPS, composed mostly of polysaccharides, and proteins \cite{FLE01}, are secreted by the bacteria themselves, which are eventually embedded within a polymeric matrix. In addition to this intrinsic polymeric matrix, development of biofilm often occurs in environments with high concentrations of extrinsic polymers, such as proteins and polysaccharides. Two mechanisms have been suggested to describe the role of polymers on the biofilm stability: polymer bridging and depletion attraction. The former occurs when a polymeric network adsorbs on the surface of multiple bacterial cells to tie them together \cite{HAR73, STR03}. The latter is an entropically-driven phenomenon that is well-known  in colloidal science \cite{EBO05, MAR06, DOR12, MAO95, LEK11}. In particular, in a polymer-rich environment, the polymer-induced attraction between colloidal particles reduces the volume excluded to the polymer and thus increases its entropy. The first theory addressing the origin of depletion attraction was proposed by Asakura and Oosawa in 1958 \cite{ASA58}. Asakura-Oosawa Theory (AOT) provides an estimation of the strength and range of the depletion interaction between spherical colloidal particles in a bath of smaller non-adsorbing spherical polymeric particles.  In this model, the polymers are considered as flexible chains in a good solvent. Under these conditions, the interaction between polymer and colloidal particle is hard, while the excluded volume interactions between polymers are neglected \cite{LEK11}. 
According to this theory, the strength of interaction is proportional to the polymer concentration, while the range of attraction is in the order of the polymer-particle diameter. Although AOT was originally  developed for colloidal spheres, other authors have extended it to the case of anisotropic colloidal particles. For instance, Savenko and Dijkstra proposed an effective interaction potential between rod-like particles in a bath of polymer particles \cite{SAV06}. This attractive potential favored configurations with parallel rodlike particles in contact.

The role of polymeric particles in the stabilization of biofilms has been identified as fundamental only very recently. For instance, Dorken \textit{et al} studied the effect of the overproduction of polysaccharides by the bacteria, observing enhanced aggregation of cells to produce biofilm-like structures \cite{DOR12}. These authors suggested that this aggregation was driven by depletion attraction. On the other hand, Secor and co-workers studied the effect of depletion attraction on  bacterial aggregation in the context of chronic infections and found that addition of external polymeric particles induced aggregation in a flagellar motility- and  biofilm-deficient ($\Delta$\textit{fliC} $\Delta$\textit{pelA} $\Delta$\textit{pslBCD} $\Delta$\textit{algD}) strain of \textit{Pseudomona aeruginosa} \cite{SEC18}. The importance of depletion attraction in biofilm formation has been mainly recognised in fluids with high  concentration of bacteria \cite{EBO05}. However, the very early stages of biofilm formation are characterized by a limited number of individual cells. In these stages, what role depletion attraction plays is still unclear and open to discussion. An additional element influencing the biofilm formation is the  ability of bacteria to diffuse on the substrate to which they attach. The expected Brownian diffusion, typical of colloidal particles, may be partially or completely inhibited in an especially crowded environment. Jiménez-Fernández \textit{et al} showed that a $\Delta$\textit{fleQ} strain of \textit{Pseudomonas putida} lacking the ability to attach irreversibly to surfaces did not form a biofilm, but exhibited Brownian motion,  \cite{JIM16}. In contrast, single cells of the wild-type \textit{P.\,putida} strain attached irreversibly to surfaces and grew into clonal biofilm microcolonies, suggesting that bacterial  ability to engage in Brownian motion is inversely related to biofilm formation.

To gain a better understanding of the impact of bacterial diffusion on the development of a biofilm, we recently developed and applied an Individual-based Model (IbM) to simulate the early stages of biofilm formation \cite{ACE18}. IbM assumes that the origin and subsequent growth of a community or population of bacteria (\textit{e.g.} biofilm) can be explained by considering the main features of individual bacteria and how they interact with each other \cite{WAN10, HOR14, KRE01, KRE02, PIC04}. In our work, we investigated the influence of individual bacterial growth and division versus bacterial diffusion on the development of biofilms. We observed that high mobility of individual bacteria -- which may be due to deficient attachment to the surface or low medium viscosity -- prevented formation of compact biofilm microcolonies, and microcolony formation was restored by increasing the medium viscosity. This molecular simulation study disregarded the presence of the polymer and only focused on the competition between bacterial diffusion and growth, which can be summarized by the following parameter

\begin{equation}\label{eq0}
\Gamma=\frac{t_{\rm dif}}{t_{\rm gr}}
\end{equation}

\noindent where $t_{\rm dif}$ is the time taken by a bacterium to diffuse a distance equal to its thickness, whereas $t_{\rm gr}$ is the time this bacterium takes to double its length. Because no polymer was included, the effect of depletion forces was clearly not considered \cite{ACE18}. Experimental observations of the microcolonies formed  by a $\Delta$\textit{fle}Q \textit{P.\,putida} strain showed a surprisingly good qualitative agreement with the simulation results. Adding polysaccharide (dextran sulfate), which is known to increase viscosity \cite{ANT12}, conferred the $\Delta$\textit{fleQ}  strain the ability to form microcolonies indistinguishable from those of the wild-type strain. However, how polymer-induced depletion forces determine the morphology of the bacterial colony and the time scales of its early-stage formation remains an open question. These forces may provide an alternative to that formulated in \cite{ACE18} explanation to experimental observations.

In the present work, we expand the IbM biofilm model by explicitly incorporating a non-adsorbing polymer, whose size and concentration can be independently varied. We notice that other authors have investigated the impact of depletion forces on the formation of biofilm in the recent past \cite{GOS15}, but these works have mostly focused on the advanced steps of the biofilm growth, where the bacterial concentration is already very large. In contrast, our goal is to assess the influence of  the presence of polymers on biofilm development from a single surface-attached cell to colonies formed by a limited number of bacteria. In this context, the effect of depletion forces, with entropic and therefore collective origin, may have to be nuanced. To ponder the validity of our simulation results, we compare them with experimental observations and find an excellent qualitative agreement. 

\section{Methods}
\subsection*{Simulation Methods}

We have modeled the influence of polymeric particles on biofilms growth with an extension of the model reported in \cite{ACE18}. More specifically, we are only modeling the first stages of the  growth of the biofilm, which can therefore be considered bidimensional. In our model, it is assumed that the bacteria only move by the effect of the interaction with other bacteria or polymer particles and by passive diffusion, having lost any possibility of active motion. The aggregation phenomena in motile bacteria has been studied by other authors in the past \cite{PER06}. In particular, a rod-like bacterial cell is modeled as a spherocylinder whose moves are restricted to two dimensions only. A spherocylinder consists of a cylinder of initial elongation $L_0$ capped by two hemispheres of diameter $\sigma$. Accordingly, the initial aspect ratio of the cells is $L^*_0 \equiv L_0/\sigma +1 $. To reproduce the dimensions of  \textit{P.\,putida}, we have chosen $L^*_0=2.6$ \cite{ACE18}. From this initial aspect ratio, the particles grow by polar elongation at constant velocity $v_{\rm gr}$ up to a maximum aspect ratio of  $L_m^*= 2L_0^*$. When the bacterium reaches this elongation, it is divided in two identical bacteria with aspect ratio $L_0^*$. All bacteria are assumed to have the same $v_{\rm gr}$ and consequently divide simultaneously. Some trials with a gaussian-distributed velocity, centered in $v_{gr}$ and standard deviation $0.1v_gr$, have also been carried out, and no significant differences have been detected. In contrast, polymer particles are modeled as spheres of diameter $\sigma_p$. Bacteria interact with each other and with the polymer via the following soft-repulsive potential \cite{PIE15, MOR18, MOR19}:

\begin{equation}\label{eq1}
U_{ij}\,= \left\{ \begin{array}{cc}
4 \epsilon_{ij} \left[ \left(\frac{1}{d^*_m}\right)^{12} -
\left( \frac{1}{d^*_m}\right)^{6} + \frac{1}{4} \right] & ~~   d^*_{m} \leq \sqrt[6]{2}
\\
0 & ~~ d^*_{m} > \sqrt[6]{2} \end{array} \right.
\end{equation}

\noindent where $i$ and $j$ are generic rod-like (bacteria) or spherical (polymer) particles, $d^*_{m}$=\,$d_{m}$/$\sigma_{ij}$ denotes the minimum distance between them and $\sigma_{ij}$ is the sum of their radii. More specifically, $U_{ij}$ mimics the steric repulsions and is the same as that applied in our previous work \cite{ACE18}.
$d_m$ is the minimum distance between segments of length L that describe the cores of the spherocylinders. For additional details on the computation of the minimum distance between two spherocylinders, being a spherical particle a particular case of a spherocylinder with elongation $L=0$, we refer the interested reader to Ref.\,\cite{VEG94}. In the present work, we have assumed the same interaction strength for both rod-rod and sphere-rod pairs, hence $\epsilon_{ij}=\epsilon$. On other hand, the non-adsorbing polymer particles are invisible to each other and can freely overlap \cite{VRI76}.

Particle movement has been modeled by Brownian dynamics (BD) simulations at constant volume. In BD simulations, the particle trajectories are obtained by integrating the Langevin equation forward in time. For polymer particles, the position $\textbf{r}_p$ of a particle $p$ changes in time according to the expression

\begin{equation}\label{eq2}
{\bf r}_p(t+\Delta t) = {\bf r}_p(t)+\frac{D_{p}}{k_BT} {\bf F}_p(t)\Delta t+ \sqrt{2D_{p} \Delta t} {\bf R_0}(t)
\end{equation}

\noindent where $D_{p}= D_0/(3\pi\sigma_p)$ is the infinite-dilution diffusion coefficient of a sphere; $D_0=D^*_0 \sigma^2/\tau$ is a diffusional parameter that depends on temperature, particle-surface adhesion energy and medium viscosity; $\tau$ is the time unit; $D^*_0=0.1$ is a constant; $\textbf{F}_p$ is the total force acting on particle $p$; and ${\bf R}_0$ a gaussian random vector of variance 1 and zero mean. Similarly, the trajectories of the center of mass of each individual bacterium, $\textbf{r}_b$, and the orientation of its longitudinal axis, $\textbf{\^u}_b$, evolve in time according to the following set of equations:

\begin{equation}
\label{eq2b}
\begin{split}
{\bf r}^{\parallel}_b(t+\Delta t) =
{\bf r}^{\parallel}_b(t)+\frac{D_{\parallel}}{k_BT} {\bf
	F}^{\parallel}_b(t)\Delta t + \\
\,\,\,\,\,\,\,\,\,\, + \sqrt{2D_{\parallel} \Delta t}
R^{\parallel} \textbf{\^{u}$_{b}$}(t)
\end{split}
\end{equation}
\begin{equation}
\begin{split}
{\bf r}^{\perp}_b(t+\Delta t) =    {\bf r}^{\perp}_b(t)+
\frac{D_{\perp}}{k_BT} {\bf F}^{\perp}_b(t)\Delta
t+ \\
\,\,\,+ \sqrt{2D_{\perp} \Delta t} \left[ R^{\perp}_1 \textbf{\^{v}$_{b,1}$}(t)+ R^{\perp}_2 \textbf{\^{v}$_{b,2}$}(t) \right] \\
\end{split}
\end{equation}
\begin{equation}
\begin{split}
\textbf{\^{u}$_{b}$}(t+\Delta t) = \textbf{\^{u}$_{b}$}(t)+
\frac{D_{\vartheta}}{k_BT} {\bf T}_b(t)\times
\textbf{\^{u}$_{b}$}(t)\Delta t+ \\
\,\,\,+ \sqrt{2D_{\vartheta} \Delta t} \left[ R^{\vartheta}_1
\textbf{\^{w}$_{b,1}$}(t)+R^{\vartheta}_2
\textbf{\^{w}$_{b,2}$}(t) \right]
\end{split}
\end{equation}

\noindent where $ \textbf{r}^{\parallel}_b$ and $\textbf{r}^{\perp}_b$ are the projections of $\textbf{r}_b$ on the directions parallel and perpendicular to $\textbf{\^{u}}_b$, respectively; $\textbf{F}^{\parallel}_b$ and $\textbf{F}^{\perp}_b$ are the parallel and perpendicular components of the total force acting on $b$ and ${\bf T}_b$ is the total torque acting over \textit{b} due to the interactions with other particles of the fluid. Details on the calculation of these forces and torques from the interaction potential of Eq. \ref{eq1} are available in  \cite{VEG90}. The Brownian dynamics of the particle is induced through a set of independent gaussian random numbers of variance 1 and zero mean: $R^{\parallel}$, $R^{\perp}_1$, $R^{\perp}_2$, $R^{\vartheta}_1$ and $R^{\vartheta}_2$, and unitary vectors perpendicular to {\bf\^{u}}$_{b}$, denoted above as $\textbf{\^{v}}_{b,m}$ and $\textbf{\^{w}}_{b,m}$ ($m$=\,1, 2). The diffusion coefficients, $D_{\parallel}$, $D_{\perp}$ and $D_{\vartheta}$ were calculated by employing the analytical expressions proposed by Shimizu for prolate spheroids \cite{SHI62}:

\begin{equation}\label{eq3}
\begin{split}
D_{\perp}&=D_{0}\frac{(2a^{2}-3b^{2})S+2a}{16\pi(a^{2}-b^{2})}b,\\
D_{\parallel}&=D_{0}\frac{(2a^{2}-b^{2})S-2a}{8\pi(a^{2}-b^{2})}b,\\
D_{\vartheta}&=3D_{0}\frac{(2a^{2}-b^{2})S-2a}{16\pi(a^{4}-b^{4})}b,
\end{split}
\end{equation}

\begin{equation}
\begin{split}
\mbox{with\,\,\,\,\,}
S=\frac{2}{\sqrt{a^{2}-b^{2}}}\log\frac{a+\sqrt{a^{2}-b^{2}}}{b},\\
(a=\,(L+\sigma)/2,\,\,\,b=\,\sigma/2)\,\,\,\,\,\,\,\,\,\,\,\,\,\,\,\,\,\,\,\,
\end{split}
\end{equation}

The time step was automatically set according to $\Delta t =\Delta t_{\rm max}\exp(-f_{\rm m}/\chi)+10^{-7}\tau$, with $f_{\rm m}$ the module of the instantaneous maximum force between any two particles in reduced units. This choice of the time step allows a relatively fast evolution of the simulation, minimizing the numerical errors associated with the integration of Langevin equations. After some trials, we have set $\Delta t_{\rm max}=10^{-3}\tau$ and $\chi=1000$.

In all the cases studied, the initial configuration consists of a single bacterium of elongation $L_0^*$ with orientation $\textbf{\^{u}}_1=(1,0)$ and $N_p$ polymer particles randomly distributed in a square simulation box of area $40 \times 40\sigma^2$. The polymer particles are initially located at a distance of least $\sigma$ from the bacterium. Because our aim is reproducing the conditions in which a biofilm grows in a bath of polymer, periodic boundary conditions and minimum-image convention have been applied \cite{FRE}. These are standard simulation techniques that allow to approximate  the properties of infinite systems using finite simulation box. To avoid interactions of a bacterial colony with its own image, we have limited the number of division to 6 and, consequently, no more than 64 bacteria are simultaneously present in the system. This is also consistent with the fact that, in their later stage of proliferation and maturation, biofilms are not two-dimensional, but rather form a multilayer structure that grows in the direction perpendicular to the substrate \cite{ACE18}. Under these conditions, our two-dimensional model would be less realistic.

As mentioned above, we intend to analyze the effect of explicitly incorporating a polymer into some of the systems studied in our previous work \cite{ACE18}. Specifically, we have considered those systems where, due to the low viscosity of the medium and/or lack of adhesion of bacteria to the substrate, compact biofilms were not produced. In Ref.\,\cite{ACE18}, the competition between lengthening and diffusion was exemplified by the parameter $\Gamma$ defined in Eq.\,\ref{eq0}. In particular, large values of $\Gamma$ imply slow diffusion of bacteria and formation of compact colonies, whereas low values of $\Gamma$ indicate fast diffusion and formation of dispersed domains of bacteria. To qualitatively reproduce the growth of colonies of $\Delta$\textit{fleQ} \textit{P.\,putida} strains observed experimentally, we set $\Gamma=1.67\cdot10^{-2}$ and $v_{\rm gr}=10^{-3}\sigma \tau^{-1}$ in all the simulations presented here. To establish the level of compactness of the growing microcolonies, in addition to visual inspection of individual configurations, we calculated the coverage profile and radius of gyration of the bacterial colony at different times. The coverage profile is defined as the fraction of surface covered at distance $r$ from the biofilm center of mass: $g(r)=A_o(r)/A(r)$. In particular, $A(r)$ is the area of an annulus centered in the colony center of mass with internal and external radius respectively equal to $r$ and $r+dr$, whereas $A_o(r)$ is the region covered by bacteria. Following this definition, $g(r)=1$ indicates total surface coverage. To obtain $g(r)$, we generated a large number of random points within the annulus ($P_{\rm tot}$) and counted those falling within the area occupied by bacteria ($P_{\rm b}$). The coverage profile is then $g(r) \approx P_{\rm b}/P_{\rm tot}$. We stress that this calculation can change with the degree of bacteria elongation and thus decided to calculate $g(r)$ at $L = 1.5L_0^*$, corresponding to the halfway point along the elongation/division process. 

In order to explore if the polymer influences the bacteria dynamics, we calculated two dynamical observables involved over the biofilm growth: the mean square displacement (MSD) and the orientational autocorrelation function. Both of them have been calculated between consecutive division steps, that is at constant number of bacteria. The mean square displacement reads

\begin{equation}\label{eq4}
\left< \Delta r^2(t) \right>_k =  \frac{1}{N_B} \left< \sum_{i=1}^{N_B} ({\bf r}_i(t)-{\bf r}_i(t_{B_k}))^2  \right >
\end{equation}

\noindent where $t_{B_k}$ is the time at which the $k^{\rm th}$ bacterial division occurs, i.e when trajectory $k$ starts, and $t_{B_k} < t < t_{B_{k+1}}$ the time at which the MSD for trajectory $k$ is calculated. In particular, for $k=1$, 2, 3, 4, 5 and 6, we calculate the MSD of $N_B=1$, 2, 4, 8, 16, 32 and 64 bacteria, respectively. Over each trajectory, the bacteria aspect ratio grows from $L^*_0$ to $L^*_m$. We also computed the evolution of the orientational autocorrelation function along each trajectory, defined as

\begin{equation}\label{eq5}
\langle E_{2}(t)\rangle_k =  \frac{1}{N_B}\left< \sum_{i=1}^{N_B} \frac{1}{2} \{ 3[\textbf{\^u}_i(t)·\textbf{\^u}_i(t_B)] -1\}   \right >
\end{equation}

\noindent This function measures the orientational correlation of a given bacterium over time. More specifically, if a bacterium diffuses and duplicates over the same direction, then its time-dependent orientational correlation is very large and $\langle E_{2}\rangle_{\rm N_B}$ tends to 1. By contrast, if its orientation significantly changes over a given duplication step, then $ \langle E_{2}\rangle_{\rm N_B}$ decays to zero. Coverage profile, MSD and orientational autocorrelation function have been averaged over at least 30 independent simulation runs for each $t$ and $N_B$.

\subsection*{Experimental Methods}

Phase contrast microscopy of biofilm microcolonies was performed on a Leica DMI4000B inverted microscope using the 40x objective and 1.6x ocular magnification. MRB52, a $\Delta$\textit{fleQ} derivative of \textit{P.\,putida} KT2442 \cite{navarrete}, was grown in LB medium \cite{sambrook} to mid-exponential phase. The culture was serially diluted in LB and 100 $\mu$l aliquots were transferred to wells of Costar 96 microtiter polystyrene plates (Corning). Attachment was allowed to proceed for 30 minutes at room temperature, planktonic cells were removed by washing twice with 150 $\mu$l LB, and then 50 $\mu$l LB containing the desired concentrations of dextran-sulfate (Dextran sulfate sodium salt from Leuconostoc spp from Sigma-Aldrich, molecular weight of $5\rm \cdot10^5 gr\cdot mol^{-1}$) were added to each well. Dextran polymer has been used in the pass by other authors as model of non-adsorbing polymer in cell aggregation studies \cite{NEU08,RAD09}. Microcolony growth was monitored by periodic visual inspection and colonies were photographed at the 64-128 cell stage (procedure modified from Ref.\,\cite{navarrete}).



\section{Results}

\begin{figure}[!t]
	\center
	\includegraphics[width =\columnwidth]{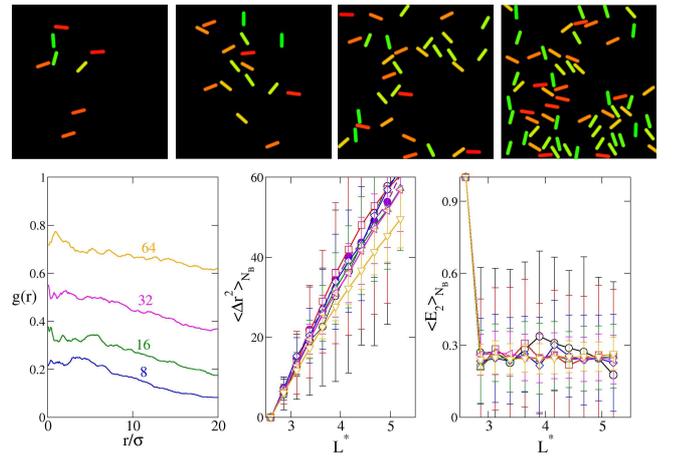}
	\caption{Top row, from left to right: snapshots of configurations in absence of polymers at $\Gamma=1.67\cdot 10^{-2}$, containing 8, 16, 32 and 64 bacteria halfway through the lengthening cycle ($L^*=3.9$). The color gradient indicates different bacterial orientations. Bottom row, left frame: coverage profile, $g(r)$, of colonies containing 8, 16, 32 and 64 bacteria at $L^*=3.9$; middle and right frames: mean-square displacement, $\rm <\Delta r^2>_{N_B}$, and orientational autocorrelation function, $\langle E_{2}(t)\rangle_{\rm N_B}$, as a function of the bacterium aspect ratio, $L^*$, in colonies containing 1 ($\CIRCLE$), 2 ($\Circle$), 4 ($\textcolor{red}{\largesquare}$), 8 ($\textcolor{Blue}\Diamond$), 16 ($\textcolor{OliveGreen}\triangle$), 32 ($\textcolor{Magenta}\medtriangleleft$) and 64 ($\textcolor{YellowOrange}\medtriangledown$) bacteria. Error bars are standard deviations.}
	\label{fig1}
\end{figure}

Early bacterial biofilm development was simulated in a $40 \times 40\sigma^2$ simulation box. The polymer particle diameter was set to $\sigma_p/\sigma = 0.05, 0.1$ and $0.5$, whereas the number of polymer particles between $N_p=0$ and $10^5$. The Stokes diameter of the dextran polymeric particles used in the experiments reported in \cite{ACE18} ($M_w=5\cdot10^5$ g mol$^{-1}$) is approximately $30\,$nm \cite{GRA58, ARM04}, equivalent to $\sigma_p=0.05\sigma$. The largest polymer concentration in \cite{ACE18} was 2.5 g L$^{-1}$. Assuming that the thickness of our simulated biofilm is $\sigma$, then the same concentration could be reproduced with approximately $N_p=10^6$ polymeric particles, a number well-above our computational capabilities. To circumvent this problem, we set the maximum number of polymer particles to $N_p=10^5$, corresponding approximately to a concentration of 1.25 g L$^{-1}$ for polymers with $\sigma_p=30\,$nm. We also have considered larger polymer sizes, \textit{de facto} expanding the range of interaction between polymers and bacteria. The early-stage of the biofilm growth in absence of polymer and at $\Gamma=1.67\cdot10^{-2}$ is shown as a reference in Fig.\,\ref{fig1} (top row). Snapshots of configurations containing 8, 16, 32 and 64 bacteria halfway through the lengthening phase ($L^*=3.9$) are displayed. This condition was also simulated in our previous work  \cite{ACE18}. Although periodic boundary conditions are employed  in the present study, there are no significant differences with our earlier study. Particles lengthen and divide on the surface and do not form biofilm-like clusters. In \cite{ACE18}, this regime, referred to as open growth, was only observed with small numbers, as further growth and division led to occupation of the inner core of the cluster. 

The qualitative insight gained from a simple visual inspection of these snapshots is consistent with the surface coverage profiles, $g(r)$, shown in the left frame of the bottom row of Fig.\,\ref{fig1}. In the absence of polymer, a similar behavior to that reported in \cite{ACE18} is observed. At $N_B \le 16$, the surface close to the colony's center of mass is scarcely covered by bacteria, and coverage decreases progressively at increasing distances from the center of mass. The fraction of covered surface increases with the number of cells, but the colony's central core is never completely covered in the sequences shown here. In addition, the slow decay of this function confirms that bacteria are dispersed over the surface and not aggregating.

\begin{figure}[!t]
	\center
	\includegraphics[width =\columnwidth]{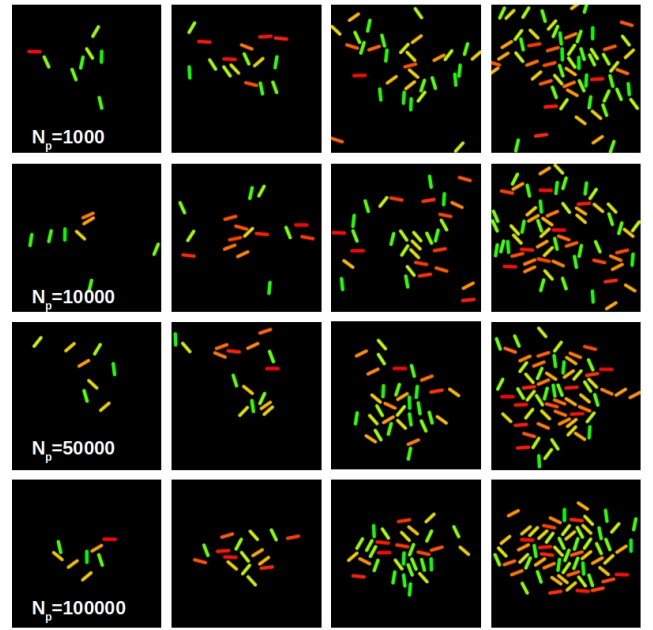}
	\caption{Clusters of 8, 16, 32 and 64 bacteria halfway through the lengthening cycle ($L^*=3.9$) containing polymer particles of diameter $\sigma_p=0.05\sigma$. From top to bottom: $N_p= 10^3, 10^4, 5\cdot 10^4$ and $10^5$. The color gradient indicates different bacterial orientations. Polymer particles are not shown.}
	\label{fig2}
\end{figure}


\begin{figure}[!t]
	\center
	\includegraphics[width =\columnwidth]{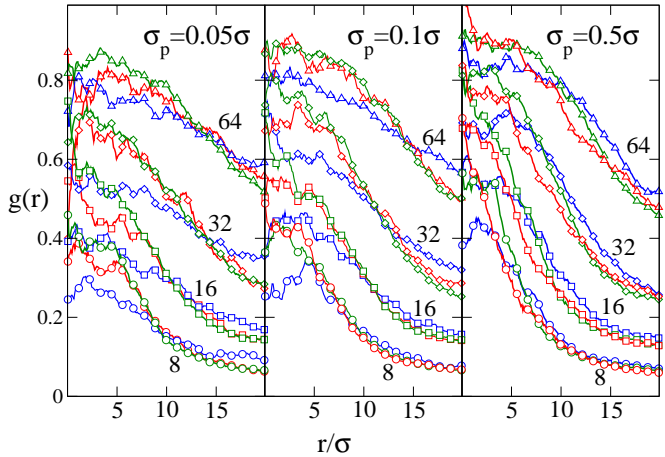}
	\caption{Surface coverage profiles $g(r)$ for colonies of 8 (circles), 16 (square), 32 (diamonds) and 64 (triangles) bacteria with $L^*=3.9$. The number of polymers is $N_p=10^4$ (blue lines and symbols), $5\cdot 10^4$ (red lines and symbols) and $10^5$ (green line and symbols). The diameter of the polymer particles is $\sigma_p=0.05\sigma$ (left frame), $\sigma_p=0.1\sigma$ (middle frame) and $\sigma_p=0.5\sigma$ right frame.}
	\label{fig3}
\end{figure}

Upon addition of a non-adsorbing polymer, the behavior of the bacterial colony starts to change, especially so at large $N_p$. This can be observed in the sequence shown in Fig.\,\ref{fig2}, where colonies containing different number of polymer particles with $\sigma_p=0.05\sigma$ are displayed. At relatively low number of polymer particles ($N_p=10^3$ and $10^4$), the colony evolves as it was unaffected by the presence of the polymer. This is confirmed by the $g(r)$ shown in the left panel of Fig.\ \ref{fig3}, which is indeed very similar to the $g(r)$ calculated at $N_p=0$, specially for low concentration of polymers. In contrast, at larger polymer concentrations, significantly more compact colonies are formed. This compaction, already appreciated at $N_p=5\cdot 10^4$, becomes more evident at $N_p=10^5$, where the trends observed in the $g(r)$ indicate that surface coverage is more pronounced in the center of the biofilm. This change in the coverage is observed for all the values of $N_B$. It should be noticed that such surface coverage is still far from being complete and the resulting colonies are still weakly packed.


\begin{figure}[!t]
	\center
	\includegraphics[width =\columnwidth]{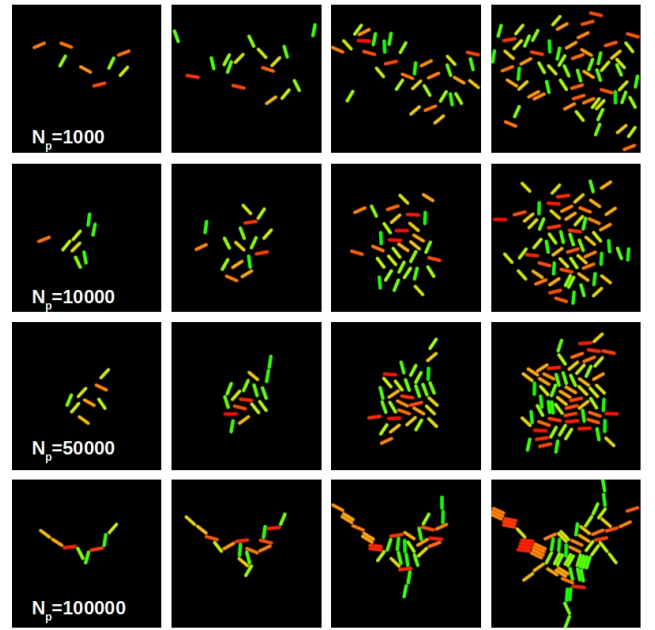}
	\caption{Colonies of 8, 16, 32 and 64 bacteria halfway through the lengthening cycle ($L^*=3.9$) containing polymer particles of diameter $\sigma_p=0.5\sigma$. From top to bottom: $N_p= 10^3, 10^4, 5\cdot 10^4$ and $10^5$. The colour gradient indicates different bacterial orientations. Polymer particles are not shown.}
	\label{fig5}
\end{figure}



At larger polymer particle diameters, the clustering of bacteria and the subsequent formation of compact colonies is more evident and offers a set of intriguing configurations at the highest polymer concentrations. With reference to Fig.\,\ref{fig5}, where $\sigma_p/\sigma=0.5$ (see supplementary information for $\sigma_p/\sigma=0.1$), we observe moderately packed bacterial colonies up to $N_p=10^4$. At larger $N_P$, these colonies become more and more compact and the scattering of bacteria on the surface is sensibly reduced. These observations are supported by the analysis of $g(r)$ in middle and right panels of Fig.\,\ref{fig3} for $\sigma_p/\sigma=0.1$ and $\sigma_p/\sigma=0.5$ respectively. In these figures, we notice an evident difference between the coverage of the colony core and its periphery, indicating preferential bacterial compaction at the core. Nevertheless, even in this case, we still observe incomplete coverage that leaves a significant portion of the surface exposed to the surroundings, in contrast to the full coverage observed at large $\Gamma$ in \cite{ACE18}. Systems with especially large polymer particles ($\sigma_p/\sigma=0.5$) show a very intriguing behavior at $N_p=10^5$ across the complete sets of cell divisions investigated here. As observed in the bottom row of Fig.\,\ref{fig5}, the bacteria produce elongated fliaments at the very early stages of growth. Such a conformation gradually changes as the biofilm grows and dendritic colonies displaying a rather compact core surrounded by filamentous extensions are formed.  The existence of such a compact core is suggested by the $g(r)$ reported in the right frame of Fig.\,\ref{fig3} for largest number of polymers, which indicates that, in the latter division stage, approximately 85\% of the biofilm core is occupied by bacteria. We would like to stress here that, while the configurations shown in Figs.\,\ref{fig1}, \ref{fig2} and \ref{fig5} report the evolution of specific, exemplary simulations, very similar morphologies were observed in all the other independent realizations simulated in this study.

\begin{figure*}[!ht]
	\center
	\includegraphics[width = 2\columnwidth]{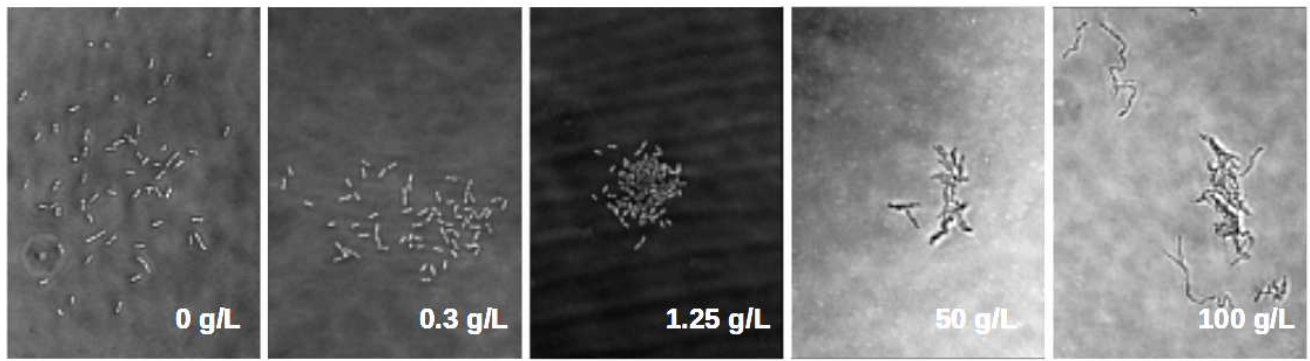}
	\caption{Micrographs of microcolonies containing $~40$ cells of $\Delta$\textit{fle}Q \textit{P.\,putida} strain MRB52 at different dextran sulfate concentration.}
	\label{figexp}
\end{figure*}

In the light of these considerations, we can conclude that our model is able to reproduce the details of the biofilm growth and its microstructure in very good qualitative agreement with the experimental observations reported in \cite{ACE18}, where the effect of adding incremental amounts of dextran sulfate to growing microcolonies of $\Delta$\textit{fle}Q \textit{P.\,putida} strain was discussed. This mutant lacks the ability to attach to surfaces and therefore diffuses faster than the wild-type strain \cite{JIM16}. Nevertheless, the addition of dextran sulfate hampers its diffusion and compact biofilms are formed \cite{ACE18}. In Fig.\,\ref{figexp}, we report micrographs of microlonies of the $\Delta$\textit{fle}Q \textit{P.\,putida} strain at different dextran sulfate concentrations, increasing up to 100 g L$^{-1}$, which is 40 times larger than the highest concentration reported in our former experiments. Depending on the polymer concentration, three main regimes were observed: (\textit{i}) the low-concentration (open-growth) regime showing relatively low-packed colonies; (\textit{ii}) the moderate-concentration regime where compact colonies, similar to those found for the wild-type strain, are observed; and (\textit{iii}) the high-concentration regime where filaments and dendritic colonies are finally found. Such a sequence is remarkably similar to the findings in our computer simulations, where the three regimes were indeed detected. However, the Stokes diameter of the dextran polymer employed in the experiments is roughly equivalent to $\sigma_p/\sigma = 0.05$, a size at which dendritic colonies were not observed (see Fig.\,\ref{fig2}). 
This is due to the fact that at $\sigma_p/\sigma = 0.05$ and $N_p=10^5$ the simulated polymer concentration is $0.125$ g L$^{-1}$, which, according to the micrographs of Fig.\,\ref{figexp}, is too low to induce the formation of dendritic colonies. More precisely, we would need to simulate approximately 40 million polymer particles of diameter $\sigma_p/\sigma = 0.05$ to see these morphologies. However, at the polymer concentration of $0.125$ g L$^{-1}$ (or $N_p=10^5$), simulations predict the formation of weakly packed colonies resembling those observed experimentally in microcolonies of $\Delta$\textit{fle}Q \textit{P.\,putida} strain at the dextran sulfate concentrations reported in Fig.\,\ref{figexp}.

In the light of these observations, we now discuss the relative impact of depletion interaction and bacteria mobility on biofilm formation. Entropy-driven depletion forces, typically observed in colloid-polymer mixtures, drive the phase separation of a colloid-rich phase from a polymer-rich phase. Depending on polymer size and concentration, the former can be highly packed and colloidal crystals can also form \cite{patti09}. In the early stages of biofilm development, depletion forces could in principle promote clustering of bacteria and thus favor the formation of compact colonies. Nevertheless, in the sequences shown above, bacteria clustering is only detected under some specific conditions. Because the presence of polymer also affects the ability of bacteria to diffuse away from their mother cell soon after division, the formation of a colony might also be determined by such reduced mobility. This is the case at low polymer concentrations, where the reduction of bacterial mobility might be more relevant than depletion forces. To address the relative importance of these two contributions, we studied the bacteria collective dynamics during their eventual aggregation into a colony. Our hypothesis is that, if the dynamics of bacteria is a collective phenomenon, then depletion interactions are expected to play a dominant role. In this case, the dynamical observables of interest, calculated during the formation of the colony, should depend on the number of bacteria involved.     

\begin{figure}[!th]
	\center
	\includegraphics[width =\columnwidth]{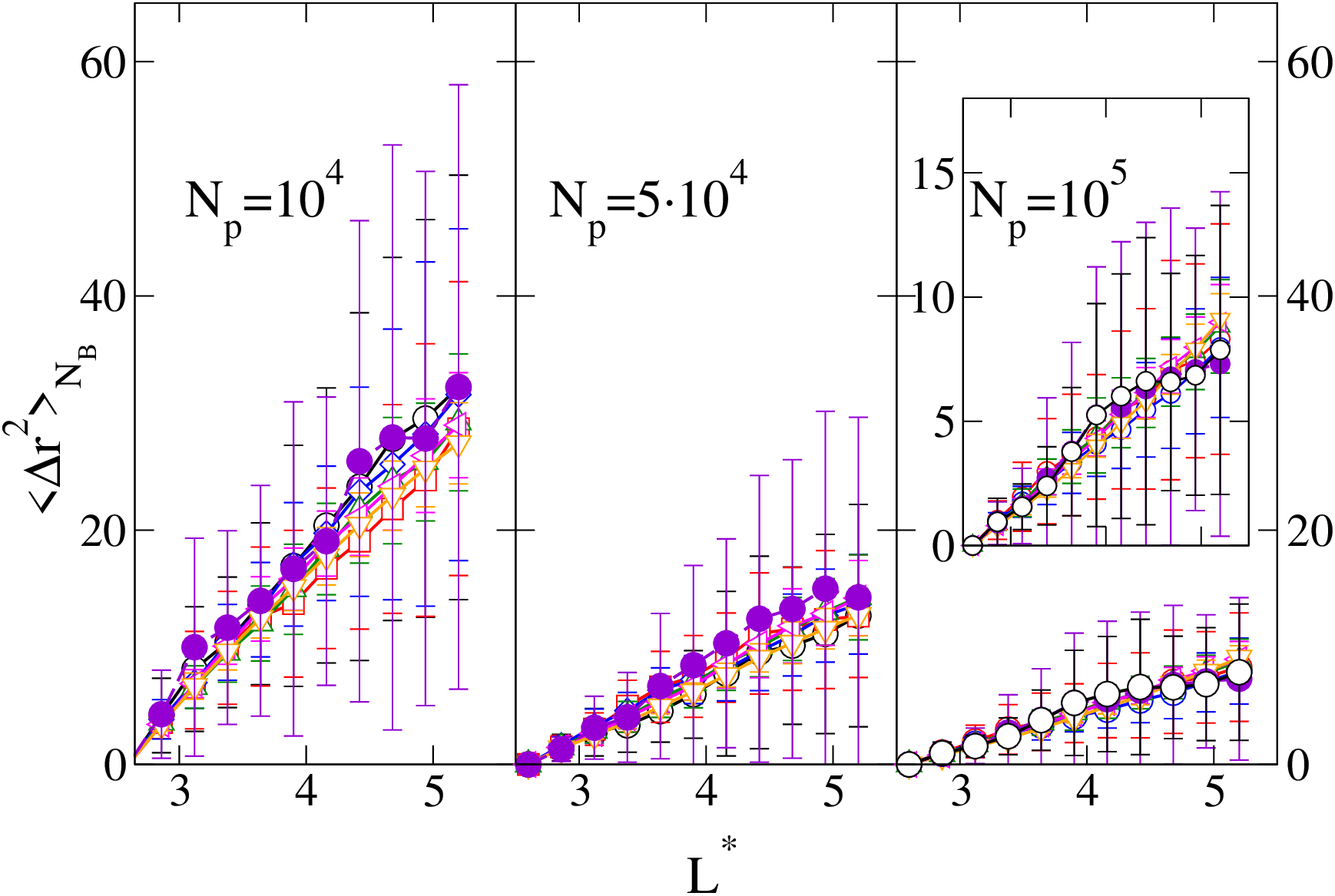}
	\caption{MSDs as a function of bacteria aspect ratio $L^*$ for colonies containing 1 to 64 bacteria and $N_p=10^4$ (left), $N_p=5\cdot 10^4$ (middle) and $N_p=10^5$ (right) polymer particles with diameter $\sigma_p=0.1\sigma$. Symbols refer to colonies containing 1 ($\textcolor{violet}\CIRCLE$), 2 ($\Circle$), 4 ($\textcolor{red}{\largesquare}$), 8 ($\textcolor{Blue}\Diamond$), 16 ($\textcolor{OliveGreen}\triangle$), 32 ($\textcolor{Magenta}\medtriangleleft$) and 64 ($\textcolor{YellowOrange}\medtriangledown$) bacteria. The inset  in the right frame magnifies the panel where it is included. Solid lines are guides for the eye and error bars represent the standard deviation of the mean.}
	\label{fig8}
\end{figure}

\begin{figure}[!ht]
	\center
	\includegraphics[width =\columnwidth]{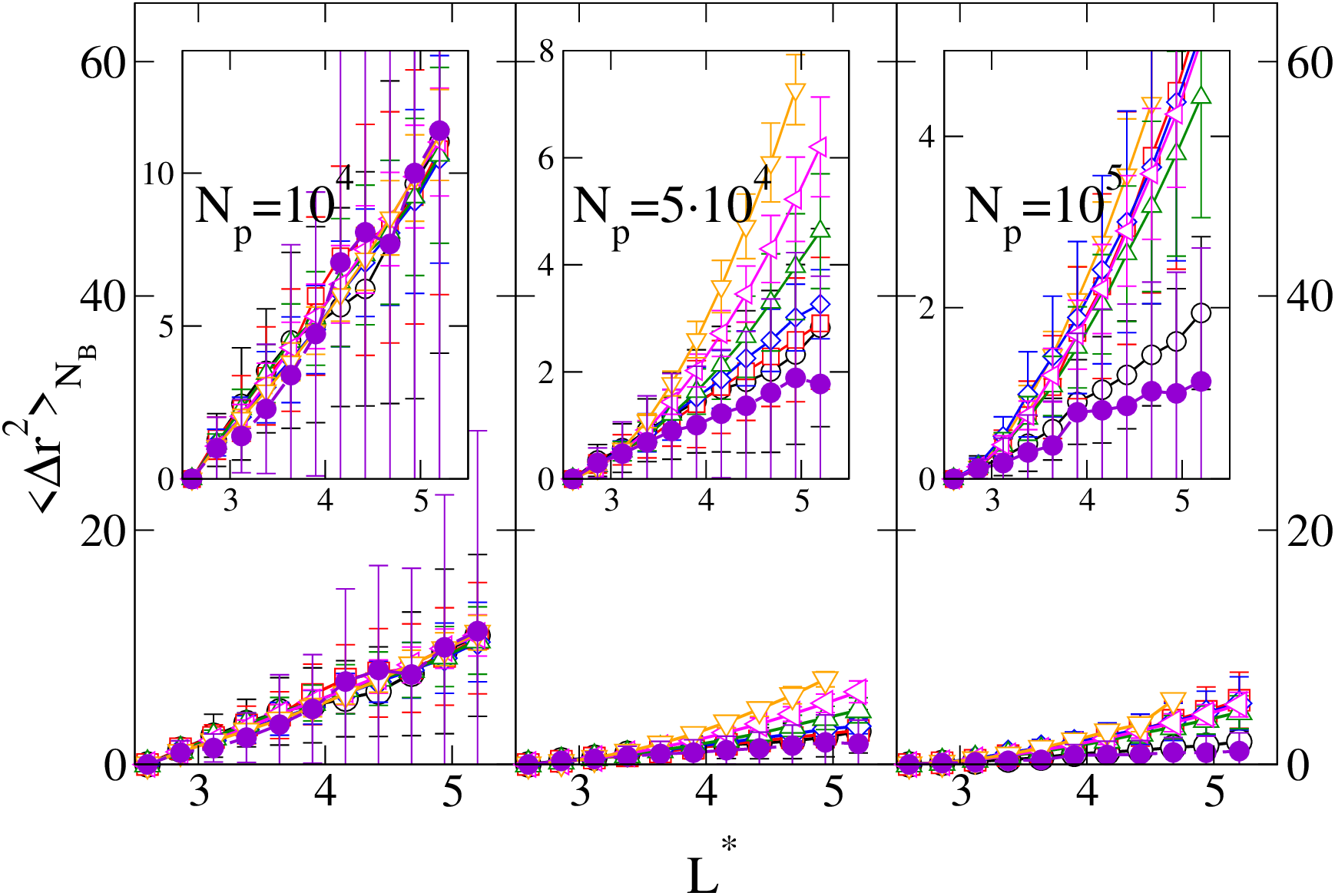}
	\caption{MSDs as a function of bacterium aspect ratio $L^*$ for colonies containing 1 to 64 bacteria and $N_p=10^4$ (left), $N_p=5\cdot 10^4$ (middle) or $N_p=10^5$ (right) polymer particles with diameter $\sigma_p=0.5\sigma$. Symbols refer to colonies containing 1 ($\textcolor{violet}\CIRCLE$), 2 ($\Circle$), 4 ($\textcolor{red}{\largesquare}$), 8 ($\textcolor{Blue}\Diamond$), 16 ($\textcolor{OliveGreen}\triangle$), 32 ($\textcolor{Magenta}\medtriangleleft$) and 64 ($\textcolor{YellowOrange}\medtriangledown$) bacteria. The inset  in the right frame magnifies the panel where it is included. Solid lines are guides for the eye and error bars represent the standard deviation of the mean.}
	\label{fig9}
\end{figure}

To this end, we calculated the MSD as a function of the bacteria aspect ratio, $L^* =L_0^*+v_{\rm gr}(t-t_B)$, for the 6 divisions (from 1 to 64 bacteria) studied here. Figs.\,\ref{fig8} and  \ref{fig9} show the so-calculated MSDs for $N_p=10^4$, $5\cdot10^4$ and $10^5$ polymer particles of size $\sigma_p/\sigma = 0.1$ and 0.5, respectively. The case $\sigma_p/\sigma = 0.05$ is available to the interested reader in the Supplementary Information. Addition of polymer particles slows down the bacterial diffusion as compared to the case where no polymer is added (see central frame in Fig.\,\ref{fig1}), especially so for larger particle diameters. Furthermore, at $\sigma_p/\sigma= 0.05$ and 0.1, the MSDs show a very weak, negligible dependence on the number of bacteria as all the curves are basically overlapping within statistical uncertainty.

As such, there is no evidence that  bacteria aggregation and colony development would be collective phenomena and we thus discard the relevance of depletion interaction and attribute a dominant role to the reduced bacterial mobility. At larger polymer particle diameters ($\sigma_p/\sigma= 0.5$), a more complex scenario is observed. At $N_P=10^4$ (left frame in Fig.\,\ref{fig9}), there is no remarkable difference with the tendencies observed for $\sigma_p/\sigma= 0.1$ as all the MSDs basically collapse on a single curve. However, at larger polymer particle numbers, we observe a clearly systematic dependence on the number of bacteria. More specifically, for $N_P=5\cdot 10^4$, bacteria of more populated colonies diffuse faster than those of smaller colonies.  Consequently, the bacteria's ability to diffuse in a polymer-rich medium depends on the number of bacteria and can thus be regarded as a depletion attraction-driven collective phenomenon. By abandoning the biofilm core to gain available volume and entropy, the polymer particles promote bacterial diffusion into this core, and more so at increasing number of bacterial cells. Therefore, the compaction observed is not necessarily due to the reduction of bacterial mobility, but more importantly to polymer-induced depletion forces. Upon further increasing the polymer concentration (right frame in Fig.\,\ref{fig9}), colonies containing at least 4 bacteria display a very similar MSD, thus excluding the occurrence of collective phenomena in established colonies. This is consistent with the filamentous structures of Fig.\,\ref{fig5}, which are not expected, because they are less thermodynamic favorable, if depletion attraction was the dominant mechanism \cite{SAV06, MAO95, LEK11}.

\begin{figure}[!t]
	\center
	\includegraphics[width =\columnwidth]{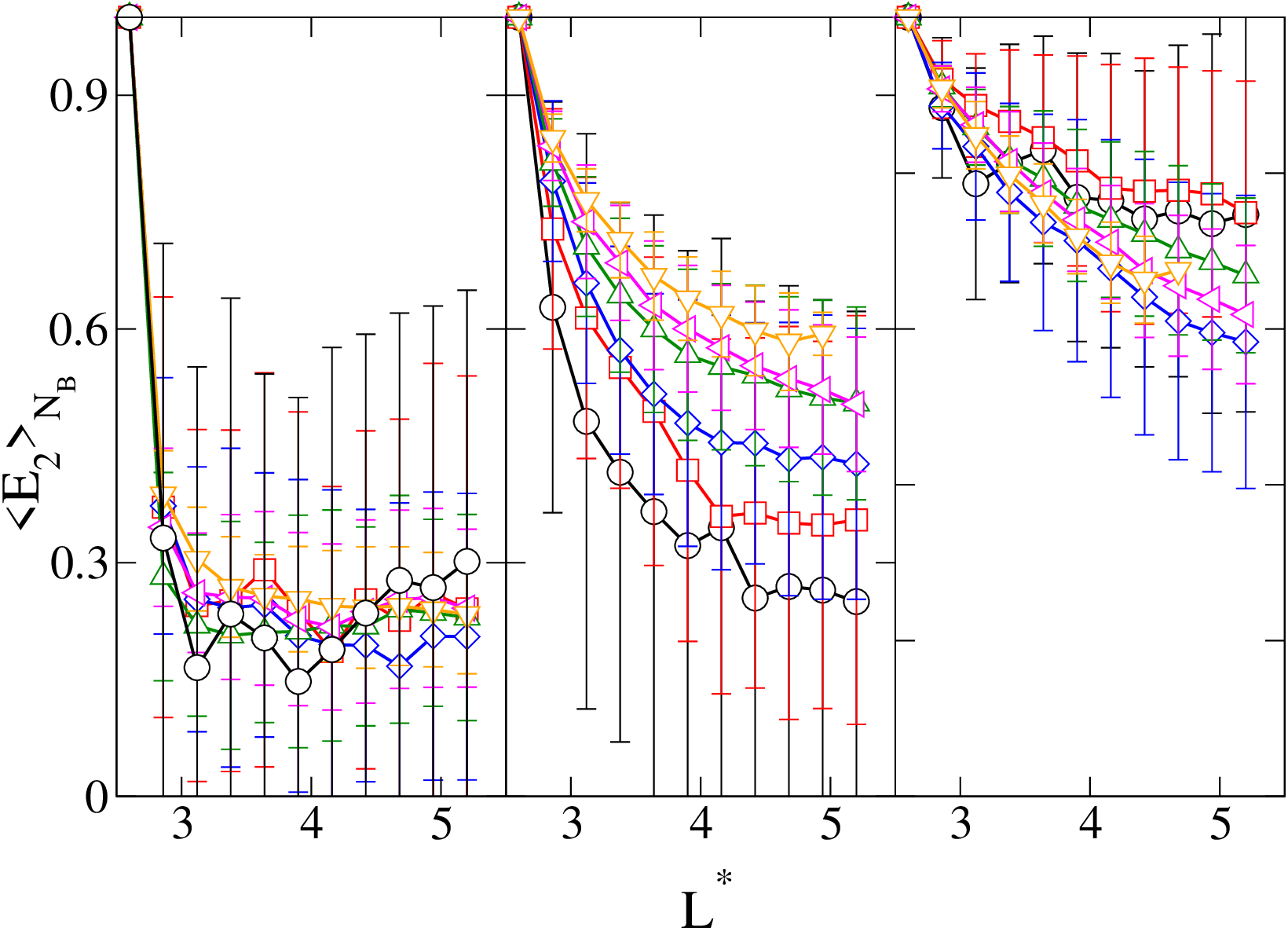}
	\caption{Orientational correlation function, $\langle E_{2}(t)\rangle_{\rm N_B}$,  as a function of bacterium aspect ratio $L^*$ for colonies containing 1 to 64 bacteria and $N_p=10^4$ (left), $N_p=5\cdot 10^4$ (middle) or $N_p=10^5$ polymer particles (diameter $\sigma_p=0.5\sigma$). Symbols refer to colonies containing 2 ($\Circle$), 4 ($\textcolor{red}{\largesquare}$), 8 ($\textcolor{Blue}\Diamond$), 16 ($\textcolor{OliveGreen}\triangle$), 32 ($\textcolor{Magenta}\medtriangleleft$) and 64 ($\textcolor{YellowOrange}\medtriangledown$) bacteria. The inset  in the right frame magnifies the panel where it is included. Solid lines are guides for the eye and error bars represent the standard deviation of the mean.}
	\label{fig10}
\end{figure}

Similar conclusions can be drawn by analysing the orientational autocorrelation functions, $\langle E_{2}(t)\rangle_{\rm N_B}$, reported in Fig.\,\ref{fig10} for the polymer particle diameter $\sigma_p/\sigma= 0.5$. At increasing number of polymer particles (from left to right frame), we observe the same tendencies found in the MSD of Fig.\,\ref{fig9}. In particular, the correlation of bacteria orientation does not seem to depend on the size of the colony at $N_p=10^4$ and  $N_p=10^6$, but shows a monotone dependence on $N_B$ at $N_p=5\cdot 10^4$. In the left frame, $\langle E_{2}(t)\rangle_{\rm N_B}$ decays approximately to 0.2, with no relevant difference when the number of bacteria increases. This behavior changes for $N_p=5\cdot 10^4$ (middle frame), where the decay of $\langle E_{2}(t)\rangle_{\rm N_B}$ slows down when the number of bacteria grows, changing from a very fast decay that suggests a full decorrelation in colonies of just 2 bacteria, to a slow decay that maintains a high degree of orientational correlation. The ejection of polymers from the biofilm core and enhanced interaction between bacteria provokes an increase in orientational correlation, a well known effect in crowded fluids of anisotropic particles \cite{CUE08,SAV06}. At $N_p=10^5$, the decay is very slow in all the cases, with a relatively large orientational correlation throughout the whole trajectory. Basically, due to the generally low mobility, each bacterium tends to maintain the orientation of its mother cell over time. 

\section{Conclusions and Final Remarks}

In summary, we have investigated how addition a non-adsorbing polymer affects the early-stage development of a biofilm of rod-like bacteria. Within the set of parameters studied, we have shown that the presence of polymer is instrumental to the clustering of bacteria in conditions in which prevalence of diffusion over growth would otherwise prevent compact colony formation. Our simulations unraveled the existence of weakly-packed, compact and dendritic colonies that closely resemble the biofilm morphologies detected experimentally in microcolonies of $\Delta$\textit{fle}Q \textit{P.\,putida} strain in the presence of dextran sulfate. We identified two polymer-induced effects that can contribute to the formation of a biofilm and to its specific morphology: depletion interaction and reduced bacteria mobility. The former has an entropic origin and typically determines phase separation of non-motile colloidal particles from a polymer; the latter limits the diffusion of a bacterium to the surroundings of its mother cell. The relevance of both effects depends on the polymer characteristic size and its concentration. For the case of small polymer particles, the observed aggregation is not caused by depletion interaction, at least in the range of concentrations studied, that is up to 0.125 g L$^{-1}$. This was inferred by investigating the dynamics of biofilm formation, which can be regarded as an individual, rather than a collective phenomenon. Consequently, the clustering of bacteria into a colony is most likely triggered by the reduction of their mobility due to the polymer particles, which keep the bacteria close to their original position after cellular division. Therefore, the colony develops because cellular division is significantly faster than bacterial diffusion. As far as larger polymer particles are concerned, the biofilm formation dramatically depends on the polymer concentration. At low polymer concentrations, the mechanism of colony growth is similar to that found for smaller polymer particles and again caused by a reduced bacteria's mobility. By contrast, at intermediate polymer concentrations, the existence of collective effects is evident and the observed aggregation is most likely due to depletion interaction. Interestingly enough, at large polymer concentrations, the collective behavior is again suppressed, but the resulting colony morphologies are very different from those observed at low concentrations as string-like and dendrite-like structures are formed. This is a signal that, under these conditions, the reduction in mobility is again dominant over the depletion interaction. Such a reduced mobility causes  new born bacteria to have a strong tendency to remain aligned over a given amount of time. The colony evolution observed in simulations at $\sigma_p/\sigma= 0.5$ was also found in experiments of a $\Delta$\textit{fle}Q \textit{P.\,putida} mutant strain in the presence of dextran sulfate. Nevertheless, the characteristic size of this polymer (close to $\sigma_p/\sigma= 0.05$) and the large polymer concentrations at which filaments and dendritic colonies are observed only allow for qualitative comparison. To account for a quantitative comparison, we should model a significantly larger number of polymer particles, in the order of $10^7$, which is currently beyond the capabilities of the present model. A valid alternative would be to introduce the effect of the polymeric particles implicitly, via an effective potential. This alternative has been successfully applied in the recent past to investigate the nucleation of colloidal crystals and liquid crystals \cite{SAV06, patti09}. Our group is now exploring the opportunity to apply this strategy to the study of biofilm development.

\begin{acknowledgements}
F.\,J. L.-C., F.\,G. and A.\,C. acknowledge the Spanish Ministerio de Ciencia, Innovaci\'on y Universidades and FEDER for funding  (project PGC2018-097151-B-I00) and C3UPO for the HPC facilities provided. A.\,P. acknowledges the Leverhulme Trust Research Project Grant No. RPG-2018-415.
\end{acknowledgements}


\end{document}